\documentclass[a4paper]{article}

\usepackage{INTERSPEECH2022}
\usepackage{cite}
\ninept

\usepackage{amsmath,amssymb}
\usepackage{physics}
\usepackage[capitalise]{cleveref}
\usepackage{xfrac}
\usepackage{csquotes}

\usepackage{pgf}
\usepackage{tikz}
\usepackage{epstopdf}
\usetikzlibrary{positioning,backgrounds,fit,dsp,calc}
\usepackage{url}
\usepackage[all]{nowidow}

\usepackage{tabularx}
\usepackage{booktabs}
\usepackage{array}
\newcolumntype{C}{ @{}>{${}}c<{{}$}@{} }

\usepackage{comment}   
\usepackage{todonotes} 
\setuptodonotes{inline}

\usepackage[acronym, shortcuts]{glossaries}
\glsdisablehyper

\newacronym{elbo}{ELBO}{evidence lower bound}
\newacronym{dft}{DFT}{discrete Fourier transform}
\newacronym{sgm}{SGM}{score-based generative model}
\newacronym{snr}{SNR}{signal-to-noise ratio}
\newacronym{gan}{GAN}{generative adversarial network}
\newacronym{vae}{VAE}{variational autoencoder}
\newacronym{ddpm}{DDPM}{denoising diffusion probabilistic model}
\newacronym{stft}{STFT}{short-time Fourier transform}
\newacronym{sde}{SDE}{stochastic differential equation}
\newacronym{dnn}{DNN}{deep neural network}
\newacronym{pesq}{PESQ}{Perceptual Evaluation of Speech Quality}

\newcommand{\xstar}[0]{x_0}
\renewcommand{\d}[0]{{\mathrm{d}}}
\newcommand{\submin}[0]{_\text{min}}
\newcommand{\submax}[0]{_\text{max}}
\newcommand{\params}[0]{{\mathbf{\theta}}}

\DeclareMathOperator{\CGauss}{\mathcal{N}_\mathbb{C}}
\DeclareMathOperator{\IdMat}{\mathbf{I}}
\newcommand{\minitimes}{{\mkern-2mu\times\mkern-2mu}}
\newcommand{\G}[1]{\textcolor{gray}{#1}}

\makeatletter
\newcommand*{\getcountref}[1]{%
\expandafter\@getcountref\csname r@#1\endcsname
}
\newcommand*{\@getcountref}[1]{%
\ifx#1\relax
0
\else
\expandafter\@car#1\@empty\@nil
\fi
}
\makeatother

\title{Speech Enhancement with Score-Based Generative Models\\ in the Complex STFT Domain}
\name{Simon Welker$^{1,2,\dagger}$, Julius Richter$^{1,\dagger}$, Timo Gerkmann$^1$\thanks{\scriptsize We acknowledge the support by DASHH (Data Science in Hamburg - HELMHOLTZ Graduate School for the Structure of Matter) with the Grant-No. HIDSS-0002 and the German Research Foundation (DFG) in the transregio project Crossmodal Learning (TRR 169).}}
\address{
  $^{1}$Signal Processing (SP), Universität Hamburg, Germany \\
  $^{2}$Center for Free-Electron Laser Science, DESY, Hamburg, Germany\\
  {\small $^{\dagger}$Authors contributed equally to this work.}}
\email{simon.welker@uni-hamburg.de, julius.richter@uni-hamburg.de, timo.gerkmann@uni-hamburg.de}

\begin{document}

\maketitle
\begin{abstract}
  \Acp{sgm} have recently shown impressive results for difficult generative tasks such as the unconditional and conditional generation of natural images and audio signals. In this work, we extend these models to the complex \ac{stft} domain, proposing a novel training task for speech enhancement using a complex-valued deep neural network. We derive this training task within the formalism of \acp{sde}, thereby enabling the use of predictor-corrector samplers. We provide alternative formulations inspired by previous publications on using generative diffusion models for speech enhancement, avoiding the need for any prior assumptions on the noise distribution and making the training task purely generative which, as we show, results in improved enhancement performance.
\end{abstract}

\section{Introduction}
Speech enhancement aims at estimating clean speech signals from audio recordings that are impacted by acoustic noise~\cite{hendriks2013dft}. The task is well studied in the signal processing literature, and conventional approaches often make assumptions regarding the statistical properties of speech signals~\cite{gerkmann2018book_chapter}. With the advent of deep learning, approaches to speech enhancement have made significant progress in the last decade. Most methods are based on a discriminative learning task that aims to minimize a certain distance between clean and noisy speech. However, since supervised methods are inevitably trained on a finite set of training data with limited model capacity for practical reasons, they may not generalize to unseen situations, e.g., different noise types, reverberation, and different \acp{snr}. In addition, some discriminative approaches have been shown to result in unpleasant speech distortions that outweigh the benefits of noise reduction~\cite{wang2019bridging}.

Instead of learning a direct mapping from noisy to clean speech, generative models aim to learn the distribution of clean speech as a prior for speech enhancement. Several approaches have utilized deep generative models for speech enhancement using \acp{gan}~\cite{pascual2017segan, baby2019sergan}, \acp{vae}~\cite{bando2018statistical,leglaive2018variance,richter2020speech,carbajal2021disentanglement,fang2021variational, bie2021unsupervised}, flow-based models~\cite{nugraha2020flow}, and more recently generative diffusion models~\cite{lu2021study, lu2022conditional}. The main principle of these approaches is to learn the inherent properties of clean speech, such as its spectral and temporal structure, which are then used as prior knowledge for making inferences about clean speech given a noisy input. Thus, they are trained solely to generate clean speech and are therefore considered more robust to different acoustic environments compared to their discriminative counterparts. In fact, generative approaches have shown to perform better under mismatched training and test conditions~\cite{fang2021variational, lu2022conditional, bando2020adaptive, bie2021unsupervised}. However, they are currently less studied and still lag behind discriminative approaches, which is a strong incentive to conduct more research to realize their full potential.

\begin{figure}[t]
\centering
\includegraphics[width=\columnwidth]{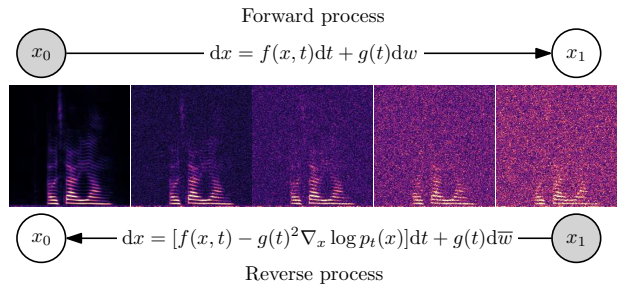}
\caption{The forward and reverse process on a spectrogram as a solution to an \ac{sde}. The reverse process gradually converts the corrupted signal $x_1$ into a clean speech spectrogram $\xstar$.}
\label{fig:sde-process}
\end{figure}

Although \acp{gan} and \acp{vae} have become a popular choice for deep generative modeling, generative diffusion models~\cite{song2019generative,ho2020denoising,song2021sde} provide a powerful alternative which has recently shown to be state-of-the-art in natural image generation~\cite{dhariwal2021diffusion}. The idea is to gradually turn data into noise, and to train a neural network that learns to invert this procedure for different noise scales (see \cref{fig:sde-process}). Other works have adopted this scheme for generating speech in the time domain using clean Mel spectrograms as a conditioner~\cite{kong2021diffwave,chen2021wavegrad}. Interestingly, the model in~\cite{kong2021diffwave} investigated its zero-shot speech denoising capabilities for different noise types. Although it was only trained to remove white noise added in the diffusion process, it already shows preliminary results for performing speech enhancement. Lu et al. built upon this idea and designed a supportive reverse process using the same architecture but with noisy spectrograms as the conditioner~\cite{lu2021study}. In their follow-up paper they devised a conditional diffusion process with an adopted forward and reverse process incorporating the noisy data into the diffusion process~\cite{lu2022conditional}. However, the derivation of their objective function is based on the assumption that the global distribution of the additive noise follows a standard white Gaussian, which is normally not the case for environmental noise. Moreover, the neural network used in their reverse process is trained to explicitly estimate the difference between clean and noisy speech, which is usually considered a discriminative learning task.

In this work, we propose an alternative diffusion process which is based on \acfp{sde} for \acfp{sgm}~\cite{song2021sde} which is a specific class of generative diffusion models (see Section~\ref{sec:background}). Using this formulation, we are able to avoid the need for any prior assumptions on the noise distribution and make the training task purely generative. Unlike~\cite{lu2021study, lu2022conditional} which work in the time domain, we perform generative speech enhancement in the complex \ac{stft} domain, i.e., working directly on the Fourier coefficients with amplitude and phase. This choice should allow deep-learning models to exploit the rich structure of speech in the time-frequency domain. In contrast to approaches that modify only the amplitudes, this choice avoids the need for phase retrieval techniques, since the phase information is enhanced directly as well. In our experiments, we demonstrate good speech enhancement performance and show that our method introduces less speech distortions compared to the baseline method.

\section{Background}
\label{sec:background}
\subsection{Score-based generative models}\label{sec:sgms}

\Acp{sgm} \cite{song2019generative, song2021sde} are generative diffusion models that center around the idea of corrupting training data with slowly increasing levels of noise (forward process), and training a \emph{score model} to reverse this corruption (reverse process) by estimating the \emph{score} $\grad_{x} \log p_\text{data}(x)$, i.e., the gradient of the log probability density with respect to the data. Once a score model is trained, an iterative procedure called Langevin dynamics can be used to draw samples from it \cite{parisi1981correlation}.


\subsection{\Acp{sgm} through stochastic differential equations}
Song et al.~\cite{song2021sde} introduced the formalism of using \acp{sde} for score-based generative modeling. Following the Itô interpretation~\cite[ch.~4]{sarkka2019sde}, the process $\{x_t\}_{t}$ can be defined by the \ac{sde}
\begin{equation}
    \d{}x_t = f(x_t, t) \d{}t + g(t) \d{}w,
\end{equation}
where $f$ is the \emph{drift}, $g$ is the \emph{diffusion}, $t$ is a coordinate which denotes how far the forward process has run, $\d{}t$ is an infinitesimal time step, and $w$ is the standard Wiener process. We stress that the \enquote{time} $t$ mentioned in this context is a conceptual idea related to the progression of the process, and is not related to the time axis of an audio signal or its \ac{stft}. According to Anderson~\cite{anderson1982reverse}, every such \ac{sde} has a corresponding \emph{reverse \ac{sde}}
\begin{equation}\label{eq:reverse-sde}
    \d{}x_t = \left[f(x_t, t) - g(t)^2 \grad_{x_t} \log p_t(x_t)\right]\d{}t + g(t)\d{}\overline{w},
\end{equation}
where $\d{}t$ is an infinitesimal negative time step and $\overline{w}$ is a standard Wiener process for time flowing in reverse. The score term $\grad_{x_t} \log p_t(x_t)$ is with respect to the log density of the process at time $t$ and can be approximated by a learned time-dependent score model $s_\params(x_t, t)$. The reverse \ac{sde} is then solved by means of some solver procedure (see \cref{sec:sampling-procedure}), providing the basis for score-based generative modeling with \acp{sde}.

\section{Proposed method}
\subsection{A stochastic process for speech enhancement}
Lu et al.~\cite{lu2022conditional} proposed the idea of linearly interpolating between clean speech $\xstar$ (at $t=0$) and noisy speech $y = \xstar+n$ (at $t=1$) along a discrete-time forward process, so that the reverse process should lead to clean speech at $t=0$. Note, that  discrete-time diffusion processes using Markov chains \cite{ho2020denoising} are different from the continuous \ac{sde} formulation used in the context of \acp{sgm} \cite{song2021sde} and in this work. Furthermore, as can be seen from their subsequent derivations, the choice of linear interpolation implies that the trained \ac{dnn} must explicitly estimate a portion of environmental noise $n$ at each step in the reverse process. From an intuitive standpoint, this occurs because it is necessary to estimate the slope between every value of $\xstar$ and $y$ to be able to formulate the reverse process. As a consequence, the resulting \ac{elbo} loss~\cite[eq.~(21)]{lu2022conditional} exhibits characteristics of a discriminative learning task, making it difficult to assess whether the network is learning a prior for clean speech per se or the mapping from noisy to clean speech.
To avoid this, we propose the following novel diffusion process based on an \ac{sde}, which leads to a \emph{pure} generative training objective without the model estimating any portion of $n$:
\begin{align}\label{eq:sde}
    \d{}x_t &= \gamma(y-x_t)\d{}t + g(t)\d{}w\\
    g(t) &= \sigma\submin \left(\sfrac{\sigma\submax}{\sigma\submin}\right)^t \sqrt{2\log\left(\sfrac{\sigma\submax}{\sigma\submin}\right)}
\end{align}
where $\sigma\submin$ and $\sigma\submax$ parameterize the variance schedule of the added Gaussian noise, and $\gamma$ is a constant that can be interpreted as a stiffness parameter for $\xstar$ being pulled towards $y$ as $t$ becomes larger. This \ac{sde} is inspired by the concept of stochastic processes that exhibit \emph{mean reversion}~\cite{chakraborty2011market}, i.e., a convergence of the mean to a particular value as $t \rightarrow \infty$. 
Since the drift $f(x_t,t) = \gamma(y-x_t)$ is affine with respect to $x_t$, the \ac{sde}~\eqref{eq:sde} describes a Gaussian process~\cite{sarkka2019sde}. Let $\IdMat$ be the identity matrix. The state distribution of this process, called perturbation kernel,
\begin{equation}\label{eq:perturbation-kernel}
    p_{0t}(x_t|\xstar,y) = \CGauss\left(x_t; \mu_{\xstar,y}(t), \sigma(t)^2 \IdMat\right),
\end{equation}
admits efficient sampling $x_t$ at an arbitrary timestep $t$ because it is fully characterized by its mean $\mu_{\xstar,y}(t)$ and variance $\sigma(t)^2$, which can be derived in closed form via eqns. (5.50, 5.53) in Särkkä \& Solin~\cite{sarkka2019sde} as:
\begin{align}
    \label{eq:mean}
    \mu_{\xstar,y}(t) &= e^{-\gamma t}\xstar + (1-e^{-\gamma t})y
    \\
    \label{eq:std}
    \sigma(t)^2 &= \frac{
        \sigma\submin^2\left(\left(\sfrac{\sigma\submax}{\sigma\submin}\right)^{2t} - e^{-2\gamma t}\right)\log(\sfrac{\sigma\submax}{\sigma\submin})
    }{\gamma+\log(\sfrac{\sigma\submax}{\sigma\submin})}
\end{align}
Since $\mu_{\xstar,y}(t)$ does not exactly reach $y$ for finite $t$, we empirically choose a stiffness $\gamma$ so that $\mathbb{E}\left[|\mu_{\xstar,y}(1) - y|^2\right]~<~10^{-3}$, where the expectation is calculated over all complex bins in a random sample of 256 spectrogram pairs from the chosen dataset. Our \ac{sde} is, in essence, a synthesis of an Ornstein-Uhlenbeck \ac{sde}~\cite{uhlenbeck1930theory} and the Variance Exploding \ac{sde} by Song et al.~\cite{song2021sde}. The affine drift term (as in an Ornstein-Uhlenbeck \ac{sde}) leads to exponential decay of the mean from $\xstar$ towards $y$, reflected in \eqref{eq:mean}, and the exponentially increasing diffusion term (as in a Variance Exploding \ac{sde}) leads to exponentially increasing corruption of features by Gaussian noise.

We note that in \eqref{eq:perturbation-kernel}, we assume circularity and a scaled identity matrix as the covariance matrix for the process. This would be a strong assumption to make if the process should model real-world additive noise. We argue that the assumption is acceptable in this case, as the noise added by the forward process is entirely artificial, intended to mask the particular characteristics of the environmental noise at $t=1$, and thus represents only a means to an end for the generative task.

\subsection{Data representation}
In this work, we represent speech signals in the complex-valued one-sided \ac{stft} domain. We therefore treat clean speech $\xstar$ and noisy speech $y$ as elements of $\mathbb{C}^{\left(\frac{F}{2}+1\right)\minitimes{}T}$, where $F$ is the \ac{dft} length and $T$ is the number of time frames, dependent on the audio length. Since the global distribution of \ac{stft} speech amplitudes is typically heavy-tailed~\cite{gerkmann2010empirical}, the information visible in untransformed spectrograms is dominated by only a small portion of bins. Furthermore, the naturally encountered amplitudes often lie far outside the interval $[0,1]$ often used in \Acp{sgm}~\cite{ho2020denoising,song2021sde}. As an engineering trick, we thus apply an amplitude transformation to all complex \ac{stft} coefficients $c$~\cite{braun2021consolidated}, in an effort to bring out frequency components with lower energy (e.g. fricative sounds of unvoiced speech) and to normalize amplitudes roughly to within $[0,1]$ to better fit the usual \ac{sgm} assumptions. The transformation and its associated inverse transformation is defined as:
\begin{equation}\label{eq:spec-transform}
    \tilde{c} = \frac{|c|^\alpha}{\beta} e^{i \angle(c)} \quad \Leftrightarrow \quad  c = \beta|\tilde{c}|^{\sfrac{1}{\alpha}} e^{i \angle(\tilde{c})}
\end{equation}
where we have chosen $\alpha=0.5$ and $\beta=3$ empirically. The forward and backward Gaussian processes, as well as the input and output of the score-based \ac{dnn} model, are then formulated and applied within this transformed domain.

\subsection{Training task}
We adopt the training strategy of \emph{denoising score matching}~\cite{vincent2011connection}, which trains a model to approximate the score $\grad_{x_t} \log p_t(x_t)$ by estimating the Gaussian noise added by the forward process at time $t$. This training strategy can be efficiently realized by sampling $t$ uniformly from $[0,1]$ and then sampling $x_t$ according to \eqref{eq:perturbation-kernel} for each data point in the training batch. The training task is to learn parameters $\params$ that minimize the following term, where $\xstar \sim p_0(x), x_t \sim p_{0t}(x_t | \xstar, y)$:
\begin{equation}\label{eq:training-task}
    \mathbb{E}_{t,\xstar,x_t|\xstar}
        \left[
            \norm{s_\params(x_t, t, y) - \grad_{x_t} \log p_{0t}(x_t | \xstar, y)}_2^2
        \right]
\end{equation}

We note that in the baseline work, CDiffuSE~\cite[eq.~(21)]{lu2022conditional}, the loss function includes a term $(y-\xstar)$, i.e., the network is in part trained to remove the environmental noise directly in each step. In contrast, our objective function in \eqref{eq:training-task} does not task the \ac{dnn} with estimating a portion of the environmental noise $n$, and $y=\xstar+n$ enters \eqref{eq:training-task} only as an input to the score-based model, functioning as a conditioning signal. Our model is thus trained to only estimate the Gaussian noise that is artificially added during the forward process. We therefore argue that the CDiffuSE training task has significant discriminative characteristics, whereas ours remains purely generative in nature.

\subsection{Speech enhancement procedure}\label{sec:sampling-procedure}
To proceed with speech enhancement, an initial complex spectrogram $x_1$ is determined by sampling from the corresponding prior distribution, with mean $y$ and standard deviation $\sigma(1)$:
\begin{equation}
    x_1 \sim \CGauss(y, \sigma(1)^2 \IdMat)
\end{equation}
We then choose a reverse sampling method to start from $x_1$ and solve the reverse \ac{sde}, from $t=1$ up until a small $t_\varepsilon \approx 0$ to avoid numerical issues close to $t=0$, see~\cite{song2021sde}. There are various possible choices of reverse sampling methods. Song et al.~\cite{song2021sde} show impressive performance of so-called \emph{Predictor-Corrector samplers}, which we therefore adopt as well. We use the specific combination of \emph{reverse diffusion sampling} as the predictor and \emph{annealed Langevin dynamics} as the corrector, as in~\cite[Algorithm~2]{song2021sde}. The corrector requires a so-called \ac{snr} parameter $r$, which we empirically set to $r=0.33$. We use one corrector step per iteration, at 50 total iterations. The enhanced audio is finally retrieved by applying the backward transformation \eqref{eq:spec-transform} and a subsequent inverse \ac{stft}.

\section{Experimental setup}
\subsection{Neural network architecture}
\begin{figure}[t]
    \centering
    \scalebox{0.87}{\hspace{-1.25cm} \newcommand{\reluoffset}{-0.7mm}
\newcommand{\drawrelu}[1]{
    \draw[line width=0.7*\dsplinewidth]
    (-1.5mm*#1,\reluoffset) -- (0,\reluoffset)
    (0,\reluoffset) -- (1.7mm*#1,\reluoffset+1.7mm*#1);
}
\newcommand{\Nvd}{0.4cm}
\newcommand{\Nodescale}{0.9}
\newcommand{\Tembscale}{\Nodescale}

\tikzstyle{normalnode}=[dspsquare, inner sep=1mm]
\tikzstyle{smallnode}=[normalnode, minimum height=0.6cm, minimum width=0.6cm, scale=\Nodescale]
\tikzstyle{split}=[dspnodefull]
\tikzstyle{convnode}=[smallnode,fill=blue!20]
\tikzstyle{normnode}=[smallnode,fill=orange!20]
\tikzstyle{actnode}=[smallnode,fill=green!20]
\tikzstyle{fcsnode}=[smallnode,fill=purple!10]
\tikzstyle{emptycirc}=[dspnodeopen, minimum size=\dspoperatordiameter]
\tikzstyle{connarrow}=[dspconn]

\begin{tikzpicture}[every node/.style={font=\footnotesize}]
    
    \node[convnode] (conv_enc) {$\mathbb{C}$Conv};
    \node[emptycirc, below=\Nvd of conv_enc] (temb_add_enc) {$+$};
    \node[normnode, below=\Nvd of temb_add_enc] (norm_enc) {$\mathbb{C}$BN};
    \node[actnode, below=\Nvd of norm_enc] (act_enc) {};
        \begin{scope}[shift=(act_enc -| act_enc)] \drawrelu{1} \end{scope}
    \node[split, right=1cm of act_enc] (enc_downsplit) {};
    \node[below=0.9cm of enc_downsplit.center, font=\footnotesize] (enc_downout) {Down$_{i+1}$};
    \node[fcsnode, right=0.3cm of conv_enc, scale=\Tembscale] (temb_fcs_enc) {$\mathbb{C}$LS};
    \node[emptycirc] at (temb_add_enc -| temb_fcs_enc) (temb_bc_enc) {$\leftrightarrow$};
    
    \node[emptycirc, right=4.25cm of act_enc] (concat_dec) {$\smallfrown$};
    \node[convnode, right=0.4cm of concat_dec] (tconv_dec) {$\mathbb{C}$Conv$^\mathsf{T}$};
    \node[emptycirc, above=\Nvd of tconv_dec] (temb_add_dec) {$+$};
    \node[normnode, above=\Nvd of temb_add_dec] (norm_dec) {$\mathbb{C}$BN};
    \node[actnode, above=\Nvd of norm_dec] (act_dec) {};
        \begin{scope}[shift=(act_dec -| act_dec)] \drawrelu{1} \end{scope}
    \node[below=0.9cm of concat_dec.center, font=\footnotesize] (dec_upin) {Up$_{i+1}$};
    \node[emptycirc] at (temb_bc_enc -| concat_dec) (temb_bc_dec) {$\leftrightarrow$};
    \node[fcsnode, scale=\Tembscale] at (temb_fcs_enc -| temb_bc_dec) (temb_fcs_dec) {$\mathbb{C}$LS};
    
    \node[above=0.5cm of conv_enc, font=\footnotesize] (input) {$(x_t, y)$ / Down$_{i}$};
    \node[above=0.5cm of act_dec, font=\footnotesize] (output) {Up$_{i}$ / $s_\theta$};
    
    \node[fcsnode, scale=\Tembscale] at ($(temb_fcs_enc)!0.5!(temb_fcs_dec)$) (temb_fcs2) {$\mathbb{C}$LS};
    \node[below=0.3cm of temb_fcs2, fcsnode, scale=\Tembscale] (temb_fcs1) {$\mathbb{C}$LS};
    \node[below=0.3cm of temb_fcs1, smallnode, fill=purple!30, scale=\Tembscale] (temb_gfp) {GFP};
    \node[below=0.3cm of temb_gfp] (t) {$t$};
    
    \node[below=.4cm of t, font=\footnotesize] (skip_label) {Skip};

    
    \draw[connarrow] (input) -- (conv_enc);
    \draw[connarrow] (conv_enc) -- (temb_add_enc);
    \draw[connarrow] (temb_add_enc) -- (norm_enc);
    \draw[connarrow] (norm_enc) -- (act_enc);
    \draw[connarrow] (temb_bc_enc) -- (temb_add_enc);
    \draw[connarrow] (temb_fcs_enc) -- (temb_bc_enc);
    \draw[connarrow,dashed] (enc_downsplit) -- (enc_downout);
    
    \draw[connarrow] (t) -- (temb_gfp);
    \draw[connarrow] (temb_gfp) -- (temb_fcs1);
    \draw[connarrow] (temb_fcs1) -- (temb_fcs2);
    \draw[connarrow] (temb_fcs2) -- (temb_fcs_enc);
    \draw[connarrow] (temb_fcs2) -- (temb_fcs_dec);
    \draw[connarrow] (temb_fcs_dec) -- (temb_bc_dec);
    
    \draw[connarrow] (act_enc) -- (concat_dec);
    \draw[connarrow] (concat_dec) -- (tconv_dec);
    \draw[connarrow] (tconv_dec) -- (temb_add_dec);
    \draw[connarrow] (temb_bc_dec) |- (temb_add_dec);
    \draw[connarrow] (temb_add_dec) -- (norm_dec);
    \draw[connarrow] (norm_dec) -- (act_dec);
    \draw[connarrow] (act_dec) -- (output);
    \draw[connarrow,dashed] (dec_upin) -- (concat_dec);

    \begin{pgfonlayer}{background}
    \node[
        behind path,
        rectangle,
        draw=gray,
        fill=white,
        fit=(conv_enc) (temb_add_enc) (norm_enc) (act_enc) (temb_bc_enc) (temb_fcs_enc),
        inner sep=2.5mm,
        ] (encoder_block) {};
    \node[anchor=north west, inner sep=3pt] at(encoder_block.south west) {\footnotesize\textcolor{gray}{Encoder}};
    
    \node[
        behind path,
        rectangle,
        draw=gray,
        fill=white,
        fit=(tconv_dec) (temb_add_dec) (norm_dec) (act_dec) (temb_bc_dec) (temb_fcs_dec),
        inner sep=2.5mm,
        ] (decoder_block) {};
    \node[anchor=north east, inner sep=3pt] at(decoder_block.south east) {\footnotesize\textcolor{gray}{Decoder}};
    \end{pgfonlayer}

    
    \newcommand{\Lsc}{0.7}
    \newcommand{\Lth}{2.5mm}
    \newcommand{\Lhd}{.5mm}
    \newcommand{\Lvd}{6mm}
    \newcommand{\Lhanc}{east}
    \newcommand{\Ltrel}{east}
    \newcommand{\Ltanc}{west}
    \newcommand{\Lcol}{gray}
    \newcommand{\Lfont}{\footnotesize\color{\Lcol}}

    \node[emptycirc, below right=1.4cm and 0.25cm of encoder_block.south west, anchor=west, scale=\Lsc]
        (legend_add) {$+$};
        \node[right=\Lhd of legend_add.\Ltrel, anchor=\Ltanc, font=\Lfont] () {Addition};
    \node[emptycirc, below=\Lvd of legend_add.\Lhanc, anchor=\Lhanc, scale=\Lsc]
        (legend_bc) {$\leftrightarrow$};
        \node[right=\Lhd of legend_bc.\Ltrel, anchor=\Ltanc, font=\Lfont] () {Broadcast};
    \node[emptycirc, below=\Lvd of legend_bc.\Lhanc, anchor=\Lhanc, scale=\Lsc]
        (legend_concat) {$\smallfrown$};
        \node[right=\Lhd of legend_concat.\Ltrel, anchor=\Ltanc, font=\Lfont] () {Channel Concat};
    \node[actnode, minimum width=6mm, minimum height=6mm, below=\Lvd of legend_concat.\Lhanc, anchor=\Lhanc, scale=\Lsc]
        (legend_crelu) {};
        \begin{scope}[shift=(legend_crelu -| legend_crelu)] \drawrelu{\Lsc} \end{scope}
        \node[right=\Lhd of legend_crelu.\Ltrel, anchor=\Ltanc, text height=\Lth, font=\Lfont] {Complex ReLU};

    \node[smallnode, fill=purple!30, right=2.9cm of legend_add.east, anchor=west, scale=\Lsc]
        (legend_gfp) {GFP};
        \node[right=\Lhd of legend_gfp.\Ltrel, text height=\Lth, anchor=\Ltanc, font=\Lfont] (legendtext_gfp) {Gaussian Fourier Projection};
    \node[fcsnode, below=\Lvd of legend_gfp.\Lhanc, anchor=\Lhanc, scale=\Lsc]
        (legend_bc) {$\mathbb{C}$LS};
        \node[right=\Lhd of legend_bc.\Ltrel, text height=\Lth, anchor=\Ltanc, font=\Lfont] {Complex Linear + Swish};
    \node[normnode, below=\Lvd of legend_bc.\Lhanc, anchor=\Lhanc, scale=\Lsc]
        (legend_norm) {$\mathbb{C}$BN};
        \node[right=\Lhd of legend_norm.\Ltrel, text height=\Lth, anchor=\Ltanc, font=\Lfont] {Complex Batch Norm};
    \node[convnode, below=\Lvd of legend_norm.\Lhanc, anchor=\Lhanc, scale=\Lsc]
        (legend_conv) {$\mathbb{C}$Conv};
        \node[right=\Lhd of legend_conv.\Ltrel, text height=\Lth, anchor=\Ltanc, font=\Lfont] {Complex Convolution};
        
    \begin{pgfonlayer}{background}
    \node[
        behind path,
        rectangle,
        draw=gray,
        dashed,
        fill=white,
        fit=(legend_add) (legend_bc) (legend_concat) (legend_crelu) (legendtext_gfp),  
        inner sep=1.5mm,
        ] (legend_block) {};
    \end{pgfonlayer}
    
\end{tikzpicture}}
    \caption{Architecture of each encoder/decoder pair in our U-Net \ac{dnn}. Activation functions and norms are applied to the real and imaginary parts separately, while the complex convolution and complex linear layers use the natural complex algebra. 
    }
    \label{fig:dnn-architecture}
\end{figure}
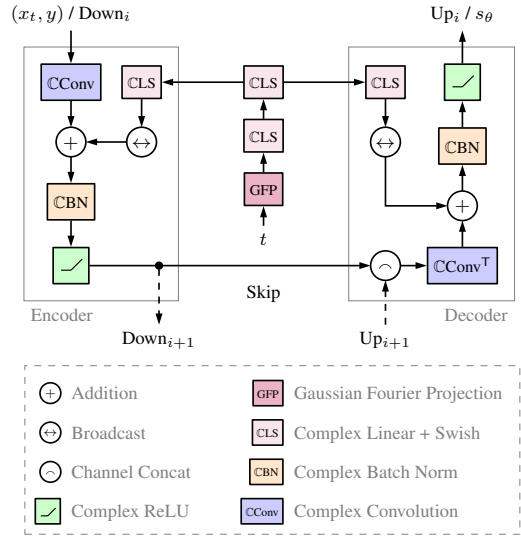

For our experiments in this paper, we construct a complex-valued U-Net architecture based on blocks shown schematically in \cref{fig:dnn-architecture} and parameterized as listed in \cref{tab:architecture-params}, resulting in 3.56M parameters overall. We adapt the \emph{Deep Complex U-Net} (DCUNet) architecture~\cite{choi2019phase} in several ways for our task: \textbf{(1)} We insert \emph{time-embedding layers} into all encoder and decoder blocks, providing the \ac{dnn} with information about the time-step $t$. We encode $t$ via 128 random Fourier feature embeddings~\cite{tancik2020fourier}, which are passed through complex-valued affine layers and activations in each block. These embeddings are broadcasted over and added to each respective channel of the layer output, as in previous \Ac{sgm}-based work~\cite{ho2020denoising,kong2021diffwave,song2021sde}. \textbf{(2)} We add multiple non-strided encoder/decoder pairs at the top of the U-Net to increase the \ac{dnn}'s capacity for finer, non-downsampled features. \textbf{(3)} We add exponentially increasing dilations in the frequency axis to the lower layers. This enlarges the receptive field in the frequency direction, to help the \ac{dnn} differentiate between spectral characteristics encountered in speech without affecting the amount of required computation. \textbf{(4)} We change the network to have one complex-valued output channel estimating the score, and two complex-valued input channels $(x_t, y)$, so that it has access to the information contained in the noisy speech $y$. \textbf{(5)} We replace all $3\minitimes3$ kernels by $4\minitimes4$ kernels, to avoid checkerboard patterns arising from the kernel size not being divisible by the stride in the transposed convolutions~\cite{odena2016deconvolution}.

\begin{table}[t]
    \centering
    \caption{Encoder and decoder parameters of our modified DCUNet architecture. $C_i$ / $C_o$ are input / output channels, $K$ is the kernel size, $S$ is the stride, and $D$ is the dilation. Tuples indicate the axes (frequency, time).}
    \label{tab:architecture-params}
    \begin{tabular}{lrrrrrr}
        \toprule
        \textbf{Depth} & \textbf{1} & \textbf{2} & \textbf{3} & \textbf{4} & \textbf{5} & \textbf{6}\\
        \midrule
        $C_i$ & 2 & 32 & 32 & 32 & 64 & 128\\
        $C_o$ & 32 & 32 & 32 & 64 & 128 & 256\\
        $K$ & (4,4) & (4,4) & (4,4) & (4,4) & (4,4) & (4,4)\\
        $S$ & (1,1) & (1,1) & (1,1) & (2,1) & (2,2) & (2,2)\\
        $D$ & (1,1) & (1,1) & (1,1) & (2,1) & (4,1) & (8,1)\\
        \bottomrule
    \end{tabular}
\end{table}

\subsection{Dataset}
We use the standardized VoiceBank-DEMAND~\cite{valentini2016investigating,thiemann2013diverse} dataset for training and testing, as was done in the baseline work (DiffuSE,~\cite{lu2021study}). We normalize the pairs of clean and noisy audio $(\xstar, y)$ by the maximum absolute value of $\xstar$. We then convert each input into the complex-valued one-sided \ac{stft} representation, using an \ac{dft} length of $F=512$, a hop length of 128 (i.e., 75\% overlap) and a periodic Hann window. We randomly crop each spectrogram to 256 \ac{stft} time frames during each epoch.

\subsection{Training procedure and hyperparameters}
We train our \ac{dnn} for 325 epochs, using the Adam optimizer~\cite{kingma2015adam} with a learning rate of $10^{-4}$ and a batch size of 32. We parameterize our \ac{sde}~\eqref{eq:sde} as follows: $\gamma = 1.5, \sigma\submin = 0.05, \sigma\submax = 0.5, t_\varepsilon = 0.03$. We track an exponential moving average of the DNN weights with a decay of 0.999, to be used for sampling~\cite{song2020improved}. We also train two baseline models~\cite{lu2021study,lu2022conditional} to compare against, using publicly available code.

\subsection{Evaluation}
For evaluation, we follow the speech enhancement procedure described in \cref{sec:sampling-procedure}. We report the scale-invariant signal-to-distortion ratio (SI-SDR), signal-to-interference ratio (SI-SIR) and signal-to-artifacts ratio (SI-SAR)~\cite{leroux2018sdr}, comparing against the noisy speech $y$ and the baseline method. We avoid the use of the \Ac{pesq} measure, since the P.862.3 standard~\cite{PESQguide} states that \enquote{there should be a minimum of 3.2s active speech in the reference}, which does not hold for the majority of audio files in the VoiceBank-DEMAND~\cite{valentini2016investigating} dataset used by us and (C)DiffuSE~\cite{lu2021study,lu2022conditional}. We provide listening examples in order to assess the perceptual quality of the compared methods\footnote{\label{web:code-and-examples}\url{https://uhh.de/inf-sp-sgmse}}.

\section{Results}

In \cref{tab:results-sisXr}, we compare the performance of our proposed method (SGMSE) with DiffuSE~\cite{lu2021study}, CDiffuSE \cite{lu2022conditional} and the noisy mixture. The results from CDiffuSE, for which executable code was not available at the time of submission, were added to the table after acceptance. We report the mean results and their 95\% confidence interval on the test set. We also report values for $m=(0.8\hat{x} + 0.2y)$, as suggested and used in~\cite{lu2021study,lu2022conditional}.

Our proposed method shows an improvement in SI-SDR of $4.6\,\text{dB}$ over DiffuSE and $3.0\,\text{dB}$ over CDiffuSE when comparing the raw model output. We can see that the three compared methods follow a trend of increasing SI-SAR and decreasing SI-SIR, with our method achieving the most favorable balance between the two metrics, resulting in the best SI-SDR. Arguably, our method also achieves more natural-sounding speech due to the lower amount of artifacts, which we confirmed by informal listening. Audio examples and code are available online\footnotemark[\getcountref{web:code-and-examples}].

\begin{figure}[t]
\centering
\includegraphics[width=\columnwidth]{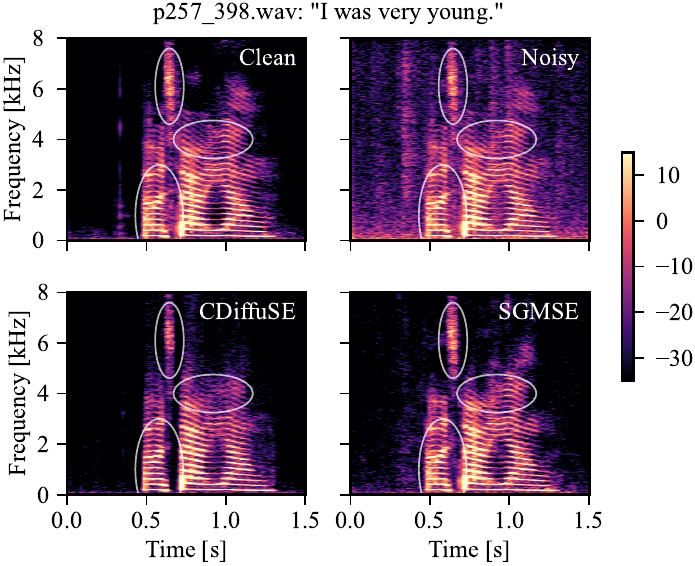}
\caption{Power spectrograms for an example audio file from VoiceBank-DEMAND. As can be seen from the highlighted regions, our approach (SGMSE) exhibits fewer voice distortions than CDiffuSE~\cite{lu2022conditional} at some expense of noise reduction.}
\label{fig:spec-comparison}
\end{figure}

This qualitative behavior is corroborated by \cref{fig:spec-comparison}, where we show the power spectrograms of an example utterance (clean and noisy) and the corresponding estimates of our method and CDiffuSE. Both methods are able to remove the environmental noise, which is clearly visible in the specified dynamic range of $50\,\text{dB}$. Furthermore, it can be seen that our method preserves the high frequencies of the fricatives better than CDiffuSE. Interestingly, CDiffuSE removes too much energy around the formants, whereas our method maintains the natural structure.

\begin{table}[t]
\centering
\caption{Average performance of our method (SGMSE) and DiffuSE~\cite{lu2021study}, comparing their raw output to the noisy speech mixture. Best values in each column are bold. Values for $m=(0.8\hat{x} + 0.2y)$, as used in~\cite{lu2021study,lu2022conditional}, are listed in gray.}
\label{tab:results-sisXr}
\begin{tabularx}{\linewidth}{X | *3{rCl}}
\toprule
\textbf{Model}& \multicolumn{3}{c}{\textbf{SI-SDR}} & \multicolumn{3}{c}{\textbf{SI-SIR}} & \multicolumn{3}{c}{\textbf{SI-SAR}}
\\
\midrule
Mixture &
    8.4 & \pm & .38 &
    8.4 & \pm & .38 &
    \multicolumn{3}{c}{$\infty$} \\
DiffuSE\textsuperscript{{\tiny\cite{lu2021study}}} &
    10.5 & \pm & .14 &
    $\mathbf{30.0}$ & \mathbf{\pm} & $\mathbf{.71}$ &
    10.8 & \pm & .11 \\
CDiffuSE\textsuperscript{{\tiny\cite{lu2022conditional}}} &
    12.1 & \pm & .10 &
    28.2 & \mathbf{\pm} & .36 &
    12.3 & \pm & .09 \\
SGMSE &
    $\mathbf{15.1}$ & \mathbf{\pm} & $\mathbf{.27}$ &
    24.9 & \pm & .42 &
    $\mathbf{15.7}$ & \mathbf{\pm} & $\mathbf{.25}$ \\
\G{DiffuSE$_m$\textsuperscript{{\tiny\cite{lu2021study}}}} &
    \G{11.4} & \G{\pm} & \G{.18} &
    \G{19.1} & \G{\pm} & \G{.46} &
    \G{13.0} & \G{\pm} & \G{.11} \\
\G{CDiffuSE$_m$\textsuperscript{{\tiny\cite{lu2022conditional}}}} &
    \G{12.6} & \G{\pm} & \G{.15} &
    \G{18.8} & \G{\pm} & \G{.34} &
    \G{14.4} & \G{\pm} & \G{.09} \\
\G{SGMSE$_m$} &
    \G{14.5} & \G{\pm} & \G{.29} &
    \G{17.9} & \G{\pm} & \G{.36} &
    \G{17.7} & \G{\pm} & \G{.25} \\
\bottomrule
\end{tabularx}
\end{table}

\section{Conclusion}
In this work, we have designed a novel stochastic process for a score-based generative modeling approach to speech enhancement in the complex \ac{stft} domain, based on the formalism of \acp{sde}. To our knowledge, this is the first work to apply \acp{sgm} in the complex time-frequency domain and, after CDiffuSE~\cite{lu2022conditional}, the second theoretically principled approach to use \acp{sgm} specifically for speech enhancement. Our approach exhibits more natural sounding reconstructions with fewer artifacts than previous methods at some expense of noise removal, which together leads to an SI-SDR improvement of about $5$ and $3\,\text{dB}$ over DiffuSE \cite{lu2021study} and CDiffuSE \cite{lu2022conditional}, respectively. Further investigation into other \acp{sde}, data representations and \ac{dnn} architectures may prove to be a fruitful research avenue for speech enhancement using generative models. An extended journal article with additional analysis and improved performance is currently in preparation \cite{richter2022journal}.

\clearpage  
\bibliographystyle{IEEEtran}
\bibliography{main}

\end{document}